\documentclass[preprint,showpacs,amsmath,amssymb,aps,prd,nofootinbib,superscriptaddress]{revtex4}
\usepackage{epsfig}
\usepackage{epstopdf}
\usepackage{color}
\begin{document}
\begin{flushright}
IBS-CTPU-16-01
\end{flushright}
\def\bea{\begin{eqnarray}}
\def\eea{\end{eqnarray}}
\def\nn{\nonumber}
\newcommand{\snu}{\tilde \nu}
\newcommand{\sll}{\tilde{l}}
\newcommand{\asnu}{\bar{\tilde \nu}}
\newcommand{\stau}{\tilde \tau}
\newcommand{\dmsnu}{{\mbox{$\Delta m_{\tilde \nu}$}}}
\newcommand{\mt}{{\mbox{$\tilde m$}}}

\renewcommand\epsilon{\varepsilon}
\def\be{\begin{eqnarray}}
\def\ee{\end{eqnarray}}
\def\lla{\left\langle}
\def\rra{\right\rangle}
\def\za{\alpha}
\def\zb{\beta}
\def\lsim{\mathrel{\raise.3ex\hbox{$<$\kern-.75em\lower1ex\hbox{$\sim$}}} }
\def\gsim{\mathrel{\raise.3ex\hbox{$>$\kern-.75em\lower1ex\hbox{$\sim$}}} }
\newcommand{\Rbs}{\mbox{${{\scriptstyle \not}{\scriptscriptstyle R}}$}}

\draft


\title{A Model for Pseudo-Dirac Neutrinos: \\
Leptogenesis
and Ultra-High Energy Neutrinos}

\textwidth 16.0cm
\textheight 24.0cm

\author{Y. H. Ahn}
\email{yhahn@ibs.re.kr}
\affiliation{Center for Theoretical Physics of the Universe, Institute for Basic Science (IBS), Daejeon, 34051, Korea}

\author{ Sin Kyu Kang}
\email{skkang@snut.ac.kr}
\affiliation{ Institute for Convergence Fundamental Study, Seoul-Tech.
       Seoul 139-743, Korea }
\author{C. S. Kim}
\email{cskim@yonsei.ac.kr}
\affiliation{ Dept. of Physics and IPAP, Yonsei University, Seoul 120-749, Korea}

\begin{abstract}
\noindent
We propose a model where sterile neutrinos are introduced to make light neutrinos  to be pseudo-Dirac particles.
It is shown how tiny mass splitting necessary for realizing pseudo-Dirac neutrinos can be achieved.
Within the model, we show how leptogenesis can be successfully generated.
Motivated by the recent observation of  very high energy neutrino events at IceCube,
we study a possibility to observe the effects of the pseudo-Dirac property of neutrinos
by performing astronomical-scale baseline experiments to uncover the oscillation effects of very tiny mass splitting.
We also discuss future prospect to observe the effects of the pseudo-Dirac property of neutrinos at high energy neutrino experiments.
\end{abstract}
 \maketitle \thispagestyle{empty}
%
\textwidth 18cm
\textheight 24.5cm

\section{Introduction}
Sterile neutrino  not only can be a good candidate for dark matter \cite{stdm} but also play an essential role in achieving
smallness of neutrino masses \cite{seesaw} and baryogenesis via leptogenesis \cite{lepto}.
The sterile neutrino states can mix with the active neutrinos and such admixtures contribute
to various processes which are forbidden in the Standard Model (SM), and  affect the interpretations of cosmological
and astrophysical observations. Thus, the masses of the sterile neutrinos and their mixing with
the active neutrinos are subject to various experimental bounds as well as cosmological and
astrophysical constraints \cite{sterile-LHC}.

While we do not have any clue to decide whether neutrinos are Dirac or Majorana particles,
here we would like to investigate for neutrinos to be pseudo-Dirac particles~\cite{Wolfenstein:1981kw}.
There have been several literatures to study neutrino as a pseudo-Dirac particle~\cite{pseudo-Dirac, redshift, beta}.
Most of them have phenomenologically studied pseudo-Dirac neutrinos with very tiny mass splitting.
In this work, we propose a model where sterile neutrinos are introduced to make light neutrinos  to be pseudo-Dirac particles.
We also show how tiny mass splitting necessary for realizing pseudo-Dirac neutrinos can be achieved.
We also examine how leptogenesis can be successfully generated within the model.

From the phenomenological point of view, one of the most important questions must be how we can probe the pseudo-Dirac neutrinos.
The magnitude of mass splittings  for pseudo-Dirac neutrinos should be smaller than  the solar, atmospheric and reactor neutrino mass scales, otherwise they
should have affected neutrino oscillations for solar, atmospheric and terrestrial neutrinos.
In order to investigate pseudo-Dirac neutrinos with very tiny mass splitting,
we need to increase the propagation length of the neutrinos, and thus
astrophysical/cosmic neutrinos detectable at neutrino telescope
can provide us with the opportunity.
In this work, we examine a possibility to observe the effects of the pseudo-Dirac property of neutrinos
by performing astronomical-scale baseline experiments to uncover the oscillation effects of very tiny mass splitting.
If the oscillation effects induced by pseudo-Dirac neutrinos with very high energy and long trajectory are prominent, then they may affect the observables detected at neutrino telescope.
The neutrino flavor composition detected from the ultra-high energy neutrino experiments
can serve as the observable to probe the effects of the pseudo-Dirac neutrinos \cite{flavor}.

Recently, IceCube experiments  announced the observation of vey high energy neutrino events \cite{IceCube}.
Analyzing the high energy neutrino events observed at IceCube, the track-to-shower ratio of
the subset with energy above 60 TeV has been studied in Ref.~\cite{Palladino:2015zua}.
They have shown that different production mechanisms for high energy neutrinos lead to
different predictions of the ratio.
Based on those results, we study how the oscillation effects induced by pseudo-Dirac neutrinos may affect the track-to-shower ratio.
Given neutrino energy and mass splittings, the oscillation effects depend on neutrino trajectory
in addition to neutrino mixing angles and CP phase.
In our numerical analysis, we take the result of global fit to neutrino data for the input of neutrino mixing angles and CP phase.
Thus, we examine how the oscillation peaks appear along with neutrino trajectory and discuss some implication on the numerical results.

This paper is organized as follows: In Sec. II, we describe a model which is an extension of the SM
through the introduction of sterile neutrinos and show how pseudo-Dirac neutrinos can be realized.
In Sec. III, we examine how leptogenesis can be successfully generated in this model.
In Sec. IV, we study how the pseudo-Dirac property of neutrinos can be probed through the results of high energy neutrino experiments.
In Sec. V, we draw our conclusions.


\section{A Model for Pseudo-Dirac Neutrinos}

In order to realize pseudo-Dirac neutrinos, let us consider
the renormalizable Lagrangian given in the charged lepton basis as
\be
-{\cal L}=\frac{1}{2}\overline{N_R^c}\,M_{R}\,N_R+ \overline{L}\,\tilde{\Phi}Y_{D}\,N_R+ \overline{L}\,\tilde{\Phi}\,Y_{DS}\,S  +
 \overline{S^c}\,\Psi\,Y_{S}\,N_R +\frac{1}{2} \overline{S^c}\,\mu\,S +h.c.~,
\label{Lag}
\ee
where $L,N_R,S$ stand for SU(2)$_L$ left-handed lepton doublet, right-handed
singlet and newly introduced singlet neutrinos, respectively, and $\tilde{\Phi}\equiv i\tau_{2}\Phi^{\ast}$.
$M_{R}$ and $\mu$ are Majorana masses for the $N_R$ and $S$ fields, respectively.
On top of the SM Higgs doublet $\Phi=(\phi^{+}, \phi^{0})^T$, an SU(2)$_L$ singlet scalar field $\Psi$ is introduced.
Assigning quantum numbers $L:1$, $N_{R}, S:1$, $\Psi:-2$, and $\Phi:0$ under the $U(1)_L$ (or $U(1)_{B-L}$) symmetry, the above Lagrangian is invariant under $U(1)_{B-L}$ when $\mu,M_R=0$. So here the parameters $\mu, M_R$ reflect soft symmetry breaking of $U(1)_L$.
When the scalar field $\Psi$  attains a vacuum expectation value (VEV), it spontaneously breaks the $U(1)_L$ (or $U(1)_{B-L}$) symmetry, but does not break the electroweak gauge symmetry. Thus its VEV is not required to lie at the electroweak scale.
%
Since the masses of Majorana neutrino $N_R$ are much larger than those of Dirac and light Majorana ones, we can integrate out the heavy Majorana neutrinos in the Lagrangian Eq. (\ref{Lag}),  resulting in the following effective Lagrangian for neutrino sectors,
\be
-{\cal L}_{\rm eff}&=& \overline{\nu_L}\,\phi^{0}\,Y_{DS}\,S -\frac{1}{2}\overline{\nu_L}\,\phi^{0}\,Y_{D}M^{-1}_{R}Y^{T}_{D}\,\phi^{0}\,\nu^c_L-
 \overline{\nu_L}\,\phi^{0}\,Y_{D}M^{-1}_{R}Y^{T}_{S}\,\Psi\,S\nonumber \\
& & -\frac{1}{2}\overline{S^c}\,\Psi\,Y_SM^{-1}_{R}Y^{T}_{S}\,\Psi\,S
 + \frac{1}{2} \overline{S^c}\,\mu\,S+h.c.~,
\ee
where $Y_{D}, Y_{S}, Y_{DS}, M_{R}$ and $\mu$ are all $3\times3$ matrices.
When the scalar fields $\Phi$ and $\Psi$ get VEVs, the mass matrix for light neutrino sector coming from the effective Lagrangian is given by
\be
{\cal M}_{\nu}=\left(\begin{array}{ccc}
 M_{\nu\nu} & M_{\nu S}  \\
 M_{\nu S}^{T} & M_{SS}  \end{array}\right),
 \label{massmatrix}
\ee
in the $(\nu^{c}_{L}, S)^T$ basis, where $M_{\nu \nu}, M_{\nu S}$ and $M_{SS}$ are $3\times 3$ matrices and given respectively by
\be
M_{\nu\nu}&=&-m_{D}M^{-1}_{R}m^{T}_{D}, \nonumber \\
M_{\nu S}&=&m_{DS}-m_{D}M^{-1}_{R}m^{T}_{S}, \nonumber \\
M_{SS} &=& \mu-m_{S}M^{-1}_{R}m^{T}_{S},
\label{massEle}
\ee
where $m_{D}=Y_{D}\langle\phi^{0}\rangle, m_{S}=Y_{S}\langle\Psi\rangle$ and $m_{DS}=Y_{DS}\langle\phi^{0}\rangle$. Note that $M_{\nu \nu}=M^T_{\nu \nu}$ and $M_{SS}=M^T_{SS}$ are the symmetric $3\times3$ Majorana left- and right-handed neutrino mass matrices, respectively.
Here we take $M_R \gg m_S \simeq m_{D}\gg \mu$, and  neutrinos become pseudo-Dirac particles when  $M_{\nu S}$ is dominant over $M_{\nu\nu}$ and
$M_{SS}$ in Eq. (\ref{massmatrix}), which reflects $m_{DS} \gg (m_D m_S)/M_R$.
Then, the mass splitting of the light neutrinos depends on the diagonal elements $M_{\nu\nu}$ and $M_{SS}$ in flavor space.
And the mixing between active states and sterile ones given as $|\tan2\theta|=|2M_{\nu S}/(M_{SS}-M_{\nu\nu})|\gg1$ becomes almost maximal.

As shown by Lim and Kobayashi~\cite{lim}, the $6\times 6$ matrix given in Eq. (\ref{massmatrix}) can be diagonalized by
\be
 W_{\nu}= X\cdot V\,,\qquad\text{with}~X=\left(\begin{array}{ccc}
 U^\ast_L & 0  \\
 0& U_R  \end{array}\right),\quad V=
\left(\begin{array}{ccc}
V_1 & i V_1 \\
V_2 & -i V_2 \end{array} \right),
\label{pDirac}
\ee
where the $3\times 3$ matrix $U_L$ corresponds to Pontecorvo-Maki-Nakagawa-Sakata (PMNS) mixing matrix, the $3 \times 3$ matrix $U_R$
is an unknown unitary matrix and $V_1$ and $V_2$ are the diagonal matrices,
$V_1=\mbox{diag}(1,1,1)/\sqrt{2}, V_2=\mbox{diag}(e^{-i\phi_1}, e^{-i\phi_2}, e^{-i\phi_3})/\sqrt{2}$
with $\phi_i$ being arbitrary phases.
The dominant matrix $M_{\nu S}$ in Eq.~(\ref{massmatrix}) can be real and positive diagonalized by biunitary transformation
\be
U^{\dag}_LM_{\nu S}U_{R}={\rm diag}(m_{1},m_{2},m_{3})\equiv\hat{M}\,.
\ee
Keeping terms up to the first order in heavy Majorana mass, the Hermitian matrix ${\cal M}^\dag_{\nu}{\cal M}_{\nu}$ can be real and positive diagonalized by a unitary transformation $W_{\nu}$ in Eq.~(\ref{pDirac});
\be
W^{\dag}_{\nu}{\cal M}^\dag_{\nu}{\cal M}_{\nu}W_{\nu} \equiv
\hat{{\cal M}^2_{\nu}}\simeq
\left(\begin{array}{ccc}
 \hat{M}^2+\hat{M}|\delta| & 0  \\
 0 & \hat{M}^2-\hat{M}|\delta|  \end{array}\right),
 \label{HermitianM}
\ee
where $\delta\equiv\hat{M}^{\ast}_{\nu\nu}+\hat{M}_{SS}$,
 and
$\hat{\cal M}_{\nu}\equiv W^{T}_{\nu}{\cal M}_{\nu}W_{\nu}={\rm diag}(m_{\nu1},m_{\nu2},m_{\nu3},m_{S1},m_{S2},m_{S3})$.
Here $m_{\nu2}=\sqrt{m^2_{\nu1}+\Delta m^2_{\rm Sol}}$ and $m_{\nu3}=\sqrt{m^2_{\nu1}+\Delta m^2_{\rm Atm}}$, with $\Delta m^2_{\rm Sol}$ and $\Delta m^2_{\rm Atm}$, respectively, being the solar and atmospheric mass-squared differences measured in neutrino oscillation experiments.
As a result, the three active neutrino states emitted by weak interactions are described in terms of the mass eigenstates as
\be
\nu_{\ell}=U_{\ell k}\frac{1}{\sqrt{2}}(\nu_k-iS^c_k),
\ee
where $\ell$ and $k$ denote  flavor and mass eigenstates, respectively, and $U\equiv U_L$ is the $3\times3$ leptonic PMNS mixing matrix. The diagonal matrix $\delta$ responsible for splitting the Dirac neutrino masses is given by
\be
\delta=\hat{\mu}-\frac{(\hat{m}_{D})^2}{\hat{M}_{R}}-\frac{(\hat{m}_{S})^2}{\hat{M}_{R}}\,,
 \label{deltaK}
\ee
in which the hat stands for a diagonalized mass matrix: $\hat{m}_D=U^{\dag}_{L}m_{D}U_{R}$, $\hat{m}_S=U^{T}_{R}m_{S}U_{R}$, $\hat{\mu}=U^{T}_{R}\mu U_{R}$, and $\hat{M}^{-1}_R=U^{\dag}_{R}M^{-1}_{R}U^\ast_{R}$.
It is easy to see that  the size of $\delta$ is very small compared with
the magnitude of $m_{D(S)}$ in the case that the lepton number violating parameter $\mu$ is small and
$m_{D(S)}\ll M_R$.
Then the mass squared difference between $m_{\nu_i}$ and $m_{S_i}$, $\Delta m^2_k (=2m_k|\delta_k|)$, can be small and thus the pairs of
the active and sterile neutrinos can form pseudo-Dirac pairs.
It is anticipated that  $\Delta m^{2}_{k}\ll\Delta m^{2}_{\rm Sol}, |\Delta m^{2}_{\rm Atm}|$,
 otherwise the effects of the pseudo-Dirac neutrinos should have been detected.
But, in the limit that $\Delta m^{2}_{k}=0$, it is hard to discern the pseudo-Dirac nature of neutrinos.
The largest $m^2_{k}$ value depends on the neutrino mass hierarchy: for normal neutrino mass hierarchy (NH), $m^{2}_{3}\gtrsim\Delta m^2_{\rm Atm}\simeq2.5\times10^{-3}\,{\rm eV}^2$ and $m^{2}_{2}\gtrsim\Delta m^2_{\rm Sol}\simeq7.5\times10^{-5}\,{\rm eV}^2$, while for inverted one (IH) $m^{2}_{2}>m^{2}_{1}\gtrsim2.5\times10^{-3}\,{\rm eV}^2$. Thus, the upper bounds for the values of $\delta_k$ are given by
 \begin{eqnarray}
 |\delta_1|\ll3.8\times10^{-5}\,\text{eV}^2/m_1\,,\qquad|\delta_2|\ll4.3\times10^{-3}\,\text{eV}\,,\qquad|\delta_3|\ll7.5\times10^{-4}\,\text{eV}\,,
 \label{Dbound1}
 \end{eqnarray}
 for NH, and for IH
 \begin{eqnarray}
 |\delta_{1,2}|\ll7.5\times10^{-4}\,\text{eV}\,,\qquad|\delta_3|\ll3.8\times10^{-5}\,\text{eV}^2/m_3\,.
 \label{Dbound2}
 \end{eqnarray}
Note that those values are very crucial for a successful low scale leptogenesis as will be discussed later.
In case of $|\hat{\mu}|\ll|\hat{m}_D^2/\hat{M}_R|$, the very tiny mass splitting between active and sterile neutrinos arises from
lepton number violating dimension-5 operators suppressed by a very high energy scale ($e.g.$ the GUT scale or Planck scale): for example, assuming normal mass hierarchy and taking $M_{R}\sim$ Planck mass$\sim1.22\times10^{19}$ GeV, then the tiny mass splittings are $\Delta m^2_{1}\ll8.6\times10^{-8}\, y_{1}^{2}\,\text{eV}^2$, $\Delta m^2_{2}\simeq8.6\times10^{-8}\, y_{2}^{2}\,\text{eV}^2$, and $\Delta m^2_{3}\simeq5\times10^{-7}\, y_{3}^{2}\,\text{eV}^2$,
where $y_{i}$ are the diagonal entries of $\hat{Y}_{D}$ or $\hat{Y}_{S}$.
On the other hand, for $|\hat{\mu}|\gg|\hat{m}_D^2/\hat{M}_R|$, the tiny mass splittings are governed by $\delta_k\simeq|\hat{\mu}_k|$.
Interestingly enough, the upper bound for $|\delta_1|$ ($|\delta_3|$) could be large enough according to the lightest neutrino mass $m_{\nu_1}\approx m_{1}$ ($m_{\nu_3}\approx m_{3}$) for NH (IH). As will be seen in Eq.~(\ref{BoundMR1}),  a successful TeV-scale leptogenesis could be viable even for a hierarchical heavy neutrino spectrum in such a way that, as the lightest neutrino mass gets lower, the corresponding scale of $\delta_1$ or $\delta_3$ increases.


\section{Leptogenesis With Pseudo-Dirac Neutrinos}

Now, let us consider how low scale leptogenesis~\footnote{See also leptogenesis in inverse seesaw neutrino models~\cite{inverse seesaw}} can be successfully generated in this scenario by
decay of the lightest right-handed Majorana neutrino before the scalar fields get vacuum expectation values. In particular, there is a new contribution to the lepton asymmetry which
is mediated by the extra singlet neutrinos.
\begin{figure}[h]
\begin{minipage}[h]{15.0cm}
\epsfig{figure=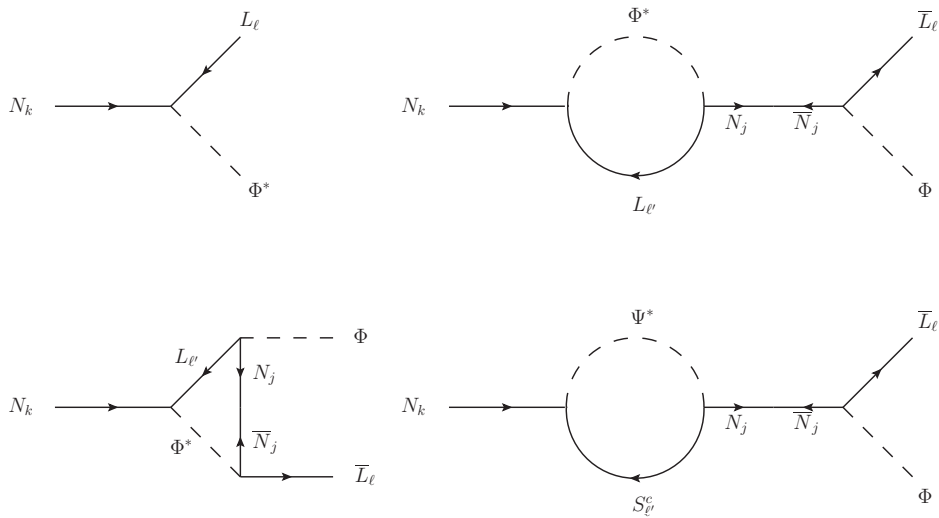,width=13.cm,angle=0}
\end{minipage}
\caption{\label{lepto} Diagrams contributing to lepton asymmetry.}
\end{figure}

Without loss of generality, we can rotate and rephase the fields to
make the mass matrices $M_{R_{ij}}$ and $\mu_{ij}$ real and diagonal. In this basis, the elements of $Y_D$ and $Y_S$ are in general  complex. As shown in Fig.~\ref{lepto}, the lepton number asymmetry from decay of the right-handed heavy neutrino into leptons and Higgs scalar required for baryogenesis is given by
\begin{eqnarray}
\epsilon_{k} &=& \sum_\ell\left[\frac{\Gamma(N_k
\to L_\ell\Phi^\ast) - \Gamma(N_k \to \overline{L}_\ell\Phi) }{\Gamma_{\rm
tot}(N_k)}\right] ,
\end{eqnarray}
where $N_k$ is the decaying right-handed neutrino and $\Gamma_{\rm tot}(N_k)$
is the total decay rate. In addition to the diagrams of the standard
leptogenesis scenario, there is a new contribution of the diagram which corresponds to the self energy correction of the vertex arisen due to the new Yukawa couplings with singlet neutrinos
and Higgs sectors. Assuming that the masses of the Higgs sectors and extra singlet neutrinos are much smaller compared to that of the right-handed neutrino, to leading order, we have
\begin{equation}
  \label{eq:vv}
\Gamma_{\rm tot}(N_k)={(Y^\dag_D Y_D+Y^\dag_S Y_S)_{kk}
\over 4\pi}M_{R_k}
\end{equation}
so that
\begin{equation}
\epsilon_k = \frac{1}{8\pi} \sum_{j\ne k} \left([ g_V(x_j)+
g_S(x_j)]{\cal T}_{kj} + g_S(x_j){\cal S}_{kj}\right),
\end{equation}
where $g_V(x_j)=\sqrt{x_j}\{1-(1+x_j) {\rm ln}[(1+x_j)/x_j]\}$,
$g_S(x_j)=\sqrt{x_j}/(1-x_j) $ with $x_j=M_{R_j}^2/M_{R_k}^2$ for
$j\ne k$,
\begin{equation}
  \label{eq:vv}
{\cal T}_{kj}={{\rm Im}[(Y^\dag_D Y_D)_{kj}^2]
 \over (Y^\dag_D Y_D +Y^\dag_S Y_S)_{kk}}\,,\qquad\qquad
{\cal S}_{kj}={{\rm Im}[(Y^\dag_D Y_D)_{kj}(Y^\dag_S
Y_S)_{kj}]
 \over (Y^\dag_D Y_D +Y^\dag_S Y_S)_{kk}}.
\end{equation}
Notice that the term proportional to ${\cal S}_{kj}$ comes from the interference of the tree-level diagram with new contribution mediated by $S$.

The newly generated B-L asymmetry is given as  $Y^{SM}_{B-L}=-\eta\,\epsilon_1 Y^{eq}_{N_1}$, where $Y^{eq}_{N_1}$ is the number density of the right-handed heavy neutrino at $T \gg M_{R_1}$ in thermal equilibrium given as $Y^{eq}_{N_1}\simeq\frac{45}{\pi^4}\frac{\zeta(3)}{g_{\ast}k_B} \frac{3}{4}$ with Boltzmann constant $k_B$ and the effective number of degree of freedom $g_{\ast}$ ($g_{\ast}=217/2$ for the SM)~\cite{Buchmuller:2005eh}. The efficient factor $\eta$ can be computed through a set of coupled Boltzmann equations which take into account processes that create or washout the asymmetry. For successful leptogenesis, the
size of the denominator of $\epsilon_1$ should be constrained by the
out-of-equilibrium condition, $\Gamma_{N_{1}} < H|_{T=M_{R_1}}$, where $\Gamma_{N_{1}}$ is the total decay width of $N_{1}$ and
$H(T=M_{R_1})=\sqrt{\frac{4\pi^3g_\ast}{45}}\frac{M^2_{R_1}}{M_{\rm Pl}}$ is the Hubble parameter at temperature $T=M_{R_1}$.
The efficiency in generating the resultant baryon asymmetry is usually controlled by the parameter defined as
\begin{eqnarray}
K\equiv\frac{\sum_\ell\Gamma(N_{1}\rightarrow L_\ell\Phi^\ast)}{H(T=M_{R_1})}=\frac{\tilde{m}_1}{m_\ast}.
\end{eqnarray}
We note that $K\ll1$ corresponds to weak washout, whereas $K\gg1$ to strong washout.
To a good approximation the efficiency factor depends on the effective
neutrino mass $\tilde{m}_1$ defined in the presence of the new Yukawa interactions with the coupling $Y_S$ by
\begin{eqnarray}
\tilde{m}_1=\frac{(Y^\dag_{D}Y_{D}+Y^{\dag}_SY_S)_{11}
}{M_{R_1}}v^2\sim2\,\delta_1,
\end{eqnarray}
which is a measure of the strength of the coupling of $N_{1}$ to the thermal bath. And the equilibrium neutrino mass is given by $m_{\ast}=\frac{16\pi^{5/2}}{3\sqrt{5}}\sqrt{g_{\ast}}\frac{v^2}{M_{\rm Pl}}\simeq1.08\times10^{-3}$ eV. Note here that the plausible range for $\tilde{m}_1$ is the one suggested by the range of the order of the $\delta_1$. In such a case, the decay rate is smaller than the expansion rate of the universe, and the particles come out of equilibrium and create a lepton asymmetry. So, the produced baryon asymmetry depends on the initial conditions in the weak washout regime. The efficiency factor for $0<\eta<1$ can be estimated by inserting this effective mass in the function~\cite{Giudice:2003jh}
\begin{eqnarray}
\eta(x)\simeq\left(\frac{3.3\times10^{-3}\,{\rm eV}}{x}+\left(\frac{x}{5.5\times10^{-4}\,{\rm eV}}\right)^{1.16}\right)^{-1}\,,
\end{eqnarray}
valid for $M_{R_1}\ll10^{14}$ GeV.
Then, the baryon-to-photon ratio results in $\eta_B\simeq-0.97\times10^{-2}\times\eta(\tilde{m}_1)\times\epsilon_1$.
From the observed one in nine year WMAP data~\cite{WMAP} $\eta^{\rm WMAP}_{B}=(6.19\pm0.14)\times10^{-10}$, we can get
the allowed range of the model parameter $\delta_i$ and some bounds on $M_{R_1}$,as will be shown later.

In a hierarchical pattern for right-handed neutrinos $M_{2,3}\gg M_{1}$, it is sufficient to consider the lepton asymmetry produced by the decay of the lightest right-handed neutrino $N_{R_1}$:
\begin{equation}
\epsilon_1 \simeq\frac{3}{16\pi} \frac{M_{R_1}}{\big(m^\dag_D m_D +m^\dag_S m_S\big)_{11}}{\rm Im}\big[\big(Y^\dag_DM_{\nu\nu} Y^\ast_D\big)_{11}\big],
\label{eps1}
\end{equation}
where $\langle\Psi\rangle\simeq\langle\phi^0\rangle=v$ is used and the loop function $g_{V}$ can be approximated as $g_{V}(x_j)=-\frac{3}{2\sqrt{x_j}}$ for $x_j\gg1$. Using the relation below Eq.~(\ref{HermitianM}), $i.e$ real and positive eigenvalues $\hat{M}_{\nu\nu}=U^{\dag}_{L}M_{\nu\nu}U^{\ast}_{L}$, we have
\begin{equation}
{\rm Im}\big[\big(\tilde{Y}^\dag_D\hat{M}_{\nu\nu} \tilde{Y}^\ast_D\big)_{11}\big]=\sum^3_{j=1} {\rm Im}\big[\big(\tilde{Y}^\dag_D\big)^2_{1j}\big]\big(\hat{M}_{\nu\nu}\big)_{j}\simeq\frac{1}{2}\sum^3_{j=1} {\rm Im}\big[\big(\tilde{Y}^\dag_D\big)^2_{1j}\big]\big|\delta_{j}\big|,
\end{equation}
where $\tilde{Y}_{D}=U^{\dag}_{L}\,Y_{D}$, and the third equality comes out from  $|\hat{\mu}|\ll|\hat{m}^2_D/M_{R}|$ with $\hat{m}_D\sim\hat{m}_S$. By letting the three-vector $\hat{Y}^{\dag}_{1j}=(\tilde{Y}^{\dag}_{D})_{1j}/\sqrt{(\tilde{Y}^\dag_{D}\tilde{Y}_{D})_{11}}$, and simply taking $(m^{\dag}_{D}m_D)_{11}\simeq(m^{\dag}_{S}m_S)_{11}$, the lepton asymmetry $\varepsilon_1$ in Eq.~(\ref{eps1}) can be simplified as
\begin{equation}
\epsilon_1 \simeq\frac{3M_{R_1}}{64v^2\pi}\sum^3_{j=1} {\rm Im}\big[\big(\hat{Y}^\dag_{1j}\big)^2\big]|\delta_j|\leq\frac{3M_{R_1}}{64v^2\pi}\,\delta_{\rm max}=\epsilon^{\rm max}_1,
\end{equation}
where $\delta_{\rm max}$ is the heaviest $|\delta_j|$. Interestingly enough, we estimate what values of $\delta_{\rm max}$ can be obtained from the solar neutrino data and GRB neutrinos
\begin{equation}
\delta_{\rm max}=\left\{
                   \begin{array}{ll}
                     \Delta m^2_1/2m_{1}\approx10^{-13}\,\text{eV}^2/2m_1\gtrsim10^{-10}\,\text{eV}, & \hbox{for NH} \\
                     \Delta m^2_3/2m_{3}\approx10^{-16}\,\text{eV}^2/2m_3, & \hbox{for IH}\\
                     \Delta m^2_k/2m_{k}\approx10^{-10}\,\text{eV}, & \hbox{for QD}
                   \end{array}
                 \right.
\end{equation}
in which $\Delta m^2_1\sim10^{-13}\,\text{eV}^2$ and $\Delta m^2_3\sim10^{-16}\,\text{eV}^2$ are taken, the QD stands for quasi-degenerate neutrino mass, and the lower bound for NH is achieved when both $m_{1}$ and $m_2$ are of the same order.
The maximal CP asymmetry $\epsilon^{\rm max}_1$ then yields the maximal baryon asymmetry $\eta^{\rm max}_{B}$ that can be produced in leptogenesis.
The lower bound on $|\varepsilon_1|$ and the upper bound on $\delta_{\rm max}$ can be used to obtain a lower bound on $M_{R_1}$
\begin{equation}
M_{R_1}\gtrsim1\,{\rm TeV}\left(\frac{0.2\,{\rm MeV}}{\delta_{\rm max}}\right)\left(\frac{|\varepsilon_1|}{10^{-7}}\right)\,,
\label{BoundMR1}
\end{equation}
which means for $\delta_{\rm max}\sim0.2$ MeV ,
the lower bound on the scale of lightest heavy neutrino will be of the order of $\gtrsim1$ TeV for a successful leptogenesis. Here the $\delta_{\rm max}\sim0.2$ MeV in the normal neutrino mass hierarchy can be obtained for $\hat{m}_{D_1}\sim0.3$ GeV and $|M_{R_1}|\sim1$ TeV
in the limit of  $|\hat{\mu}_1|\ll|(\hat{m}_{D_1})^2/M_{R_1}|$. One can obtain the $\delta_{\rm max}$ for IH in similar way.
On the other hand, in the case of QD, {\it i.e.} $m_{\nu_1}\approx m_{\nu_2}\approx m_{\nu_3}$, the lower bound on the scale of lightest heavy neutrino will be located around $M_{R}\sim$ Planck scale for a successful leptogenesis.
In its simplest scenario, thermal leptogenesis, since the baryon asymmetry is produced during the radiation dominated era, the lower bound on $M_{R_1}\gtrsim1$ TeV with $\delta_{\rm max}\sim0.2$ MeV for a hierarchical neutrino mass translates into a lower bound on the reheating temperature after inflation.

On the other hand, in a case $\delta_{\rm max}\sim2\times10^{-4}\,{\rm eV}$,
 the lower bounds on the scale of lightest heavy neutrino will be of the order of $M_R\gtrsim10^{12}$ GeV for a successful leptogenesis, which translates into a lower bound on the reheating temperature after inflation\footnote{Such a large reheating temperature is potentially in conflict with bigbang nucleosynthesis (BBN) in supersymmetric models, where upper bound on the reheating temperature as low as $10^{6}$ GeV unless $m_{3/2}\gg1$ TeV have been obtained in supergravity (SUGRA) models~\cite{Kawasaki:2004qu}.}. In such a case, the lower bound on the reheating temperature can be relaxed by considering quasi-degenerate heavy Majorana neutrinos ($M_{R_1}\simeq M_{R_2}$)~\cite{Pilaftsis:1997jf}.
So in its form, thermal production of $N_{1}$ does not need too high reheating temperature and
the Universe would not encounter the gravitino overproduction~\cite{Khlopov:1984pf, Kawasaki:2004qu}.
In the following we will see this is the case.
As shown in \cite{skk1},  the new contributions to $\epsilon_1$ could be important for the case of  $M_{R_1}\simeq M_{R_2} < M_{R_3}$ for which the asymmetry 
is approximately given by
 \begin{eqnarray}
 \epsilon_1 & \simeq & \frac{M_{R_2}}{16\pi}\frac{
           {\rm Im}\big[\big(Y^{\dag}_{D}M_{\nu\nu}Y^{\ast}_{D}\big)_{11}\big]
        -{\rm Im}\big[\big(Y^\dag_{D}\big(m_{D}M^{-1}_{R}m^T_{S}\big)Y^{\ast}_{S}\big)_{11}\big]}{\big(m^\dag_{D} m_{D}+m^\dag_S m_S\big)_{11}}\,R~,
                   \label{epsilon2}
 \end{eqnarray}
where $R$ is a resonance factor defined by $R \equiv
|M_{R_1}|/(|M_{R_2}|-|M_{R_1}|)$. In the above equation the denominator can be expressed as
 \begin{eqnarray}
 &&\sum_j\left({\rm Im}\big[(\tilde{Y}^{\dag}_{D}\big)^2_{1j}\big]\big(\hat{M}_{\nu\nu}\big)_{j}+{\rm Im}\big[\big(\tilde{Y}^{\dag}_{D}\big)_{1j}\big(\tilde{Y}^{\dag}_{S}\big)_{1j}\big]\big\{\hat{M}_j-\left(\hat{m}_{DS}\right)_j\big\}\right)\nonumber\\
 &&\simeq\sum_j\big\{{\rm Im}\big[\big(\tilde{Y}^{\dag}_{D}\big)^2_{1j}+\big(\tilde{Y}^\dag_{D}\big)_{1j}\big(\tilde{Y}^\dag_{S}\big)_{1j}\big]\big\}\frac{|\delta_{j}|}{2}
 \end{eqnarray}
where $\tilde{Y}_S=U^{T}_{R}\,Y_S$, and the second equality comes out from $|\hat{\mu}|\ll|\hat{m}^2_D/M_{R}|$ with $\hat{m}_D\sim\hat{m}_S$ leading to $\hat{M}_j-\left(\hat{m}_{DS}\right)_j\simeq|\delta_{j}|/2$ as well as $\big(\hat{M}_{\nu\nu}\big)_{j}\simeq|\delta_{j}|/2$. So, we obtain
 \begin{eqnarray}
 \epsilon_1 & \simeq & \frac{M_{R_2}}{32v^2\pi}\frac{
           \sum_j\big\{{\rm Im}\big[\big(\tilde{Y}^{\dag}_{D}\big)^2_{1j}+\big(\tilde{Y}^\dag_{D}\big)_{1j}\big(\tilde{Y}^\dag_{S}\big)_{1j}\big]\big\}|\delta_{j}|}{\big(\tilde{Y}^\dag_{D} \tilde{Y}_{D}+\tilde{Y}^\dag_S \tilde{Y}_S\big)_{11}}\,R
           \leq \frac{M_{R_2}}{64v^2\pi}\,\delta_{\rm max}\,R
 \end{eqnarray}
Similar to the hierarchical case, the lower bound on $M_{R_2}$ can be obtained by using both the upper bound on $|\varepsilon_1|$ and the lower bound on $\delta_{\rm max}$,
 \begin{eqnarray}
 M_{R_2}\gtrsim3\times10^{12-n}\,{\rm GeV}\left(\frac{2\times10^{-4}\,{\rm eV}}{\delta_{\rm max}}\right) \left(\frac{10^{n}}{R}\right) \left(\frac{|\epsilon_1|}{10^{-7}}\right).
 \end{eqnarray}
This lower bound on $M_{R_2}$ further implies a lower bound of the reheating temperature after inflation, since the abundance of gravitinos is proportional to the reheating temperature. The degree of degeneracy between two heavy neutrinos $R=10^{6-9}$ is required to achieve a successful leptogenesis, corresponding to the lower bound on $M_{R_2}\gtrsim3\times10^{3-6}$ GeV for $\delta_{\rm max}\sim2\times10^{-4}\,{\rm eV}$.


\section{Probing Pseudo-Dirac Neutrinos at Astronomical-scale  Experiments}

Now, let us consider how one can probe the effects of the pseudo-Dirac neutrinos.
A possible way to probe the pseudo-Dirac neutrinos is  to perform astronomical-scale baseline experiments to uncover the oscillation effects of very tiny
mass splitting $\Delta m^2_k$.
With the help of the mixing matrix Eq.~(\ref{pDirac}), the flavor conversion probability between the active neutrinos follows from the time evolution of the state $\nu_k$ as,
\begin{eqnarray}
 P_{\ell\ell'}\equiv P_{\nu_{\ell}\rightarrow\nu_{\ell'}}(W_{\nu},L,E)=\left|\left(W^{\ast}_{\nu}e^{-i\frac{\hat{{\cal M}}^{2}_{\nu}}{2E}L}W^T_{\nu}\right)_{\ell\ell'}\right|^2=\frac{1}{4}\left|\sum^3_{k=1}U_{\ell' k}\left\{e^{i\frac{m^2_{\nu k}L}{2E}}+e^{i\frac{m^2_{S k}L}{2E}}\right\}U^{\ast}_{\ell k}\right|^2\,,
\end{eqnarray}
where $W_{\nu}$ is the mixing matrix with which the weak gauge eigenstates, $\nu_{\ell}$, with flavor $\ell=e,\mu,\tau$ are composed of  the mass eigenstates with definite masses, $n_{k}=(\nu_k\,\,S^c_k)^T~(k=1,2,3)$, giiven as $|\nu_{\ell}\rangle=\sum^{N_{\nu}=3}_kW^{\ast}_{\ell k}|n_{k}\rangle$.


Neutrinos arriving at neutrino telescopes from astrophysical sources such as Gamma Ray Bursts (GRBs)~\cite{GRBs}, active galactic nuclei~\cite{Becker:2007sv}, and type Ib/c supernova~\cite{Kappes:2006fg} travel large distances over $\sim100$ Mpc.
Neutrino telescope observes neutrinos from extragalactic sources 
located
 a few Gpc away from the earth and  with
neutrino energy $10^5\,{\rm GeV}\lesssim E\lesssim10^7$ GeV. It has been shown~\cite{Lunardini:2000swa} that inside the GRB sources $\int V_{C,N}dt\ll1$ where the effective potentials due to the matter effects are $V_C=\sqrt{2}G_Fn_e$ with $n_e$ being the electron number density in matter and $V_N=-\sqrt{2}G_Fn_n/2$ with $n_n$ being the neutron number density in matter, so the matter effects inside the source are not relevant for neutrino oscillation, while inside the earth for $V_{C,N}\gg\Delta m^2_{k}/2E$ again the matter effect will not be significant because of the very tiny effective mixing angle.
So, we only consider neutrino oscillation in vacuum for astrophysical neutrinos.
Given neutrino trajectory $L$ and energy $E$, the oscillation effects  become prominent
when  $\Delta m^2_k\sim E/4\pi L$, 
where $L\equiv L(z)$ is a distance-measure with redshift $z$
given by \cite{redshift}
\begin{eqnarray}
 L(z)\equiv D_H\int^z_0\frac{dz'}{(1+z')^2\sqrt{\Omega_m(1+z')^3+\Omega_\Lambda}}\,,
 \label{}
\end{eqnarray}
where the Hubble length $D_H=c/H_0\simeq4.42$ Gpc with the present Hubble expansion rate $H_0=67.8\pm0.9 {\rm km}\,s^{-1}{\rm Mpc}^{-1}$~\cite{flanck}, the matter density of the Universe $\Omega_m=0.306\pm0.007$, and the dark energy density of the Universe $\Omega_\Lambda=0.694\pm0.007$~\cite{pdg}.
The asymptotic value of $L(z)$ is about $2.1$ Gpc achieved by large value of $z$, which means that  the smallest $\Delta m^2_k$ that can
be probed with astrophysical neutrinos with $E$ is $10^{-17}~\mbox{eV}^2 ~(E/\rm PeV)$ \cite{redshift}.
%
Thus, astrophysical neutrinos with  $L\simeq1$ Gpc (the flight length) and energy $E\simeq1\,{\rm PeV}$ would be useful
to probe the pseudo-Dirac property of neutrinos with very tiny mass splitting.
In this case, to observe the oscillation effects, the oscillation lengths should not be  much larger than the flight length
before arriving at neutrino telescopes in earth, that is,
\begin{eqnarray}
 L^k_{\rm osc}\simeq\left(\frac{0.8\times10^{-16}\,{\rm eV}^2}{\Delta m^2_{k}}\right)\left(\frac{E}{10^6{\rm GeV}}\right)\text{Gpc}\lesssim\text{Gpc}
\label{osc_length}
\end{eqnarray}
which means that neutrino oscillations can be measurable only when $\delta_k\gtrsim0.8\times10^{-15}$ eV.
From Eq.~(\ref{osc_length}), we see that given the tiny mass splittings $\Delta m^2_k=10^{-16 \sim -17}{\rm eV}^2$ with the energies around TeV--PeV,
 a new oscillation curve at neutrino trajectory ${\cal O}(1)$ Gpc is naively expected to occur.
Since $E/\Delta m^{2}_{k}\sim L(z) \gg E/\Delta m^{2}_{\rm Atm}$, the probability of the oscillation $\nu_\mu\rightarrow\nu_{\mu}$ over the distance $L$ is given approximately by
\begin{eqnarray}
 &P(\nu_{\mu}\rightarrow\nu_{\mu})\simeq1
 -\frac{1}{2}c^{4}_{23}\left\{4s^{2}_{23}+\sin^{2}2\theta_{12}\,c^2_{23}+\sin2\theta_{23}\sin4\theta_{12}\cos\delta_{CP}\,s_{13}\right\}\nonumber\\
 &-s^4_{23}\sin^2\left(\frac{\Delta m^{2}_{3}L}{4E}\right)-c^{4}_{23}\left\{s^4_{12}\sin^2\left(\frac{\Delta m^{2}_{2}L}{4E}\right)+c^4_{12}\sin^2\left(\frac{\Delta m^{2}_{1}L}{4E}\right)\right\}\nonumber\\
 &-\sin2\theta_{23}\sin2\theta_{12}\cos\delta_{CP}\,s_{13}\,c^{2}_{23}\left\{c^2_{12}\sin^2\left(\frac{\Delta m^{2}_{2}L}{4E}\right)-s^2_{12}\sin^2\left(\frac{\Delta m^{2}_{1}L}{4E}\right)\right\},
\end{eqnarray}
where we have ignored the terms proportional to $\sin^n\theta_{13}$ with $n\geq2$.
For the numerical analysis, we take the results
from global fit of three-flavor oscillation parameters at 1 $\sigma$ C.L.~\cite{Gonzalez-Garcia:2015qrr},
which are given in Table ~\ref{tab1}.
\begin{table}
\caption{Global fit of three-flavor neutrino oscillation parameters at 1$\sigma$.}
\centering
\begin{tabular}{c||c|c|c|c}
\hline
   mass hierarchy &$ \theta_{23} (^{\circ})$ & $\theta_{12} (^{\circ})$ &$\theta_{13} (^{\circ})$ &$\delta_{CP} (^{\circ})$ \\
   \hline  \hline
   normal &$ 42.3^{+3.0}_{-1.6}$ & $33.48^{+0.78}_{-0.75}$ &$ 8.50^{+0.20}_{-0.21}$ &$ 306^{+39}_{-70}$ \\
\hline
inverted & $49.5^{+1.5}_{-2.2}$ & $33.48^{+0.78}_{-0.75}$ & $8.51^{+0.20}_{-0.21}$ &$ 254^{+63}_{-62} $\\
\hline
\end{tabular}
\label{tab1}
\end{table}

Recently, authors in Ref.~\cite{Palladino:2015zua} analyzed the high energy neutrino events observed by IceCube, aiming to probe the initial flavor of cosmic neutrinos.
The expected number of events produced by an isotropic neutrino and antineutrino with flavor $\ell$ is given by
\begin{eqnarray}
  N=4\pi T\int dE\,\Phi_\ell(E)\,A_\ell(E)\,,
\label{}
\end{eqnarray}
 where $T$ is the time of observation, $A_\ell(E)$ is the detector effective areas, and $\Phi_\ell(E)$ is the energy dependent isotropic flux of neutrinos and antineutrinos~\cite{Palladino:2015zua}.
Then the track-to-shower ratio for the number of shower $N_S$ and track events $N_T$  in the IceCube detector\footnote{ We note that much larger detectors than the present IceCube would be required to get fully meaningful result for the test of our model in detail.}
can be expressed in terms of tiny mass splittings $\Delta m^2_{k}$, flight length $L$, neutrino mixing angles and CP phase ($\theta_{12},\theta_{23},\theta_{13},\delta_{CP}$), and initial flavor composition $\phi^0_{\ell'}$
\begin{eqnarray}
 \frac{N_T}{N_S}&=& \frac{a_\mu\,p_T\,\tilde{F}_\mu}{a_e\,\tilde{F}_e+a_\mu\,(1-p_T)\,\tilde{F}_\mu+a_\tau\,\tilde{F}_\tau}\,,
 \label{}
\end{eqnarray}
where
\begin{eqnarray}
 &\tilde{F}_\ell=\sum_{\ell'k}|U_{\ell k}|^2|U_{\ell'k}|^2\,\phi^0_{\ell'}\,,\nonumber\\
 &a_\ell=4\pi\int dE\cos^2\left(\frac{\Delta m^2_{k}L}{4E}\right)E^{-\alpha}A_\ell(E)\,,
 \label{}
\end{eqnarray}
with a spectral index $\alpha$.
Here $p_T$ is the probability that an observed event produced by a muon neutrino is a track event, which is mildly dependent on energy and approximately equals to $0.8$~\cite{Aartsen:2013jdh}.
Then above equation can be simplified to
\begin{eqnarray}
 \frac{N_T}{N_S}=\frac{\phi_\mu}{\frac{a_e}{a_\mu\,p_T}+\left(\frac{a_\tau}{a_\mu\,p_T}-\frac{a_e}{a_\mu\,p_T}\right)\,\phi_\tau+\left(\frac{1-p_T}{p_T}-\frac{a_e}{a_\mu\,p_T}\right)\,\phi_{\mu}},
\label{NTS}
\end{eqnarray}
where $\phi_{e}=1-\phi_\mu-\phi_\tau$ with $\phi_\ell\equiv\tilde{F}_\ell/(\tilde{F}_e+\tilde{F}_\mu+\tilde{F}_\tau)$ is assumed. By using the high energy neutrino events in the IceCube detector which lie in energies between $60$ TeV and $3$ PeV~\cite{IceCube, Palladino:2015zua}, Eq.~(\ref{NTS}) shows directly that track-to-shower ratio $N_T/N_S$ can give a new oscillation curve as a signal dependent on neutrino flight length if the neutrino mixing angles and CP phase, initial flavor composition, and tiny mass splittings are given as inputs.

In the limit of large or null mass splitting $\Delta m^2_{k}$, there is no oscillation effects and thus the value of $N_T/N_S$ becomes constant for a given data set of neutrino mixing angles and CP phase.
However, in the case that the oscillation effects are prominent, the value of $N_T/N_S$ can be enhanced
due to the new oscillatory term which depends on neutrino flight length, and small mass splittings.
Thus, it is possible to probe the pseudo-Dirac property of neutrinos by measuring the deviation of
$N_T/N_S$ from the expectation without the oscillation arisen  due to the tiny mass splitting.
To see how large the value of $N_T/N_S$ can be deviated by the oscillatory terms, we perform numerical
analysis by taking the values of the neutrino mixing angles and CP phase from the global fit results at $1\sigma$ level~\cite{Gonzalez-Garcia:2015qrr} as shown in Table \ref{tab1}.
We expect that the different values of $\theta_{23}$ and $\delta_{CP}$ at $1\sigma$ level for normal and inverted mass orderings provide different predictions of the track-to-shower ratio, while normal and inverted mass orderings could not be distinguished with the data at $3\sigma$ level~\cite{Gonzalez-Garcia:2015qrr}.
For the tiny mass splittings, we consider two cases:
(i) all equivalent, $\Delta m^2_{1,2,3}=\Delta m^2_{k}$, and (ii) hierarchical $\Delta m^2_{i}\gg\Delta m^2_{j}$ for ($i>j$).

\begin{figure}[h]
\begin{minipage}[h]{7.5cm}
\epsfig{figure=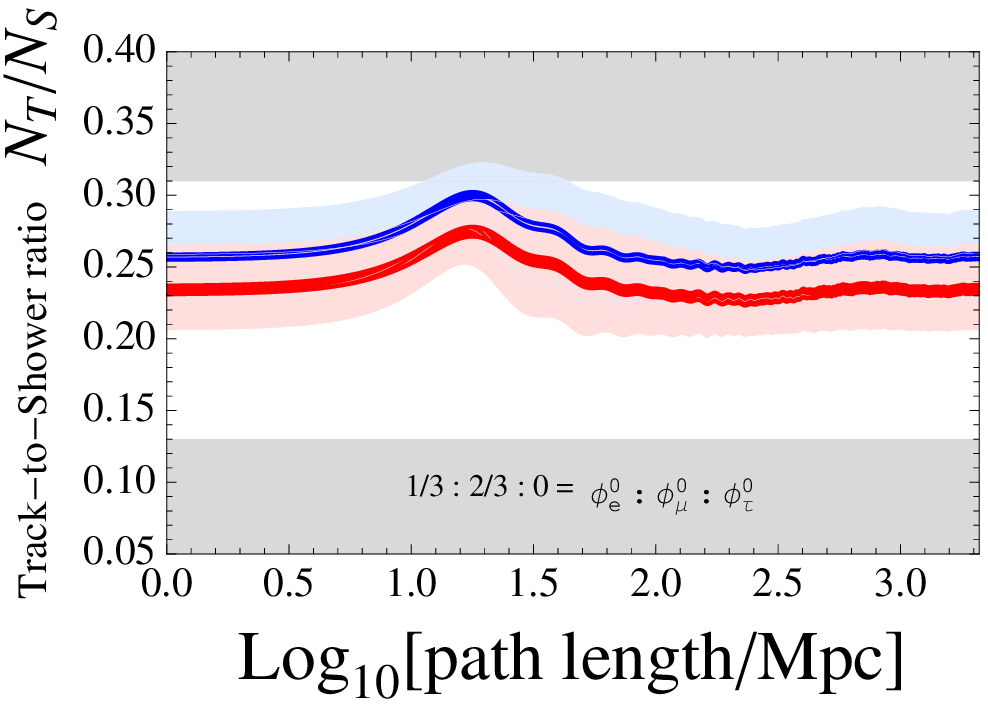,width=8.5cm,angle=0}
\end{minipage}
\hspace*{1.0cm}
\begin{minipage}[h]{7.5cm}
\epsfig{figure=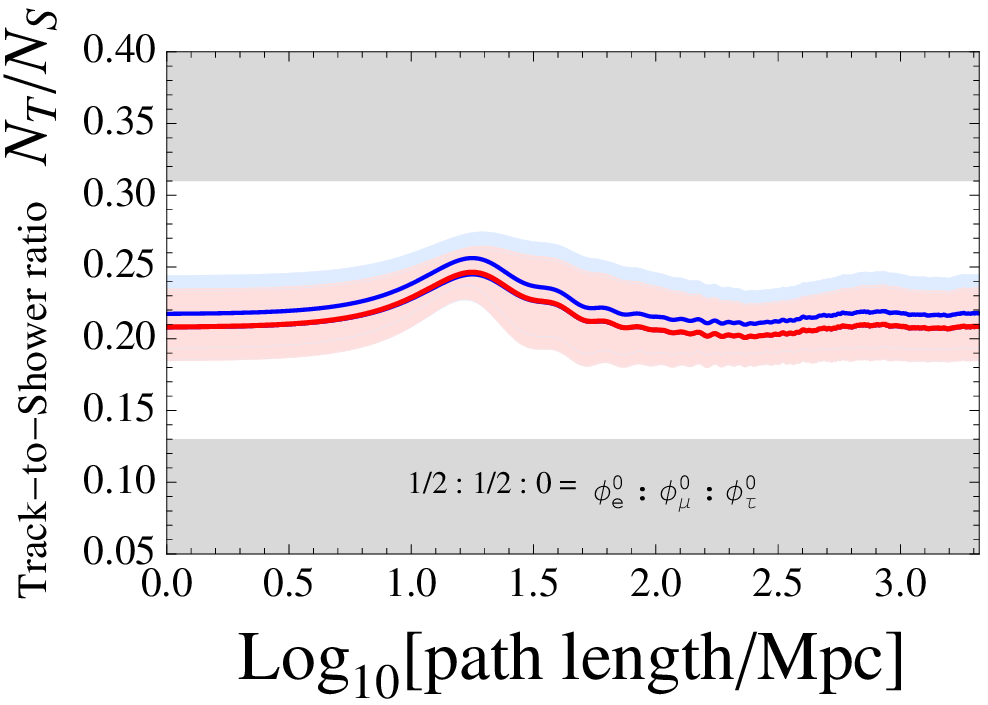,width=8.5cm,angle=0}
\end{minipage}\\
\begin{minipage}[h]{7.5cm}
\epsfig{figure=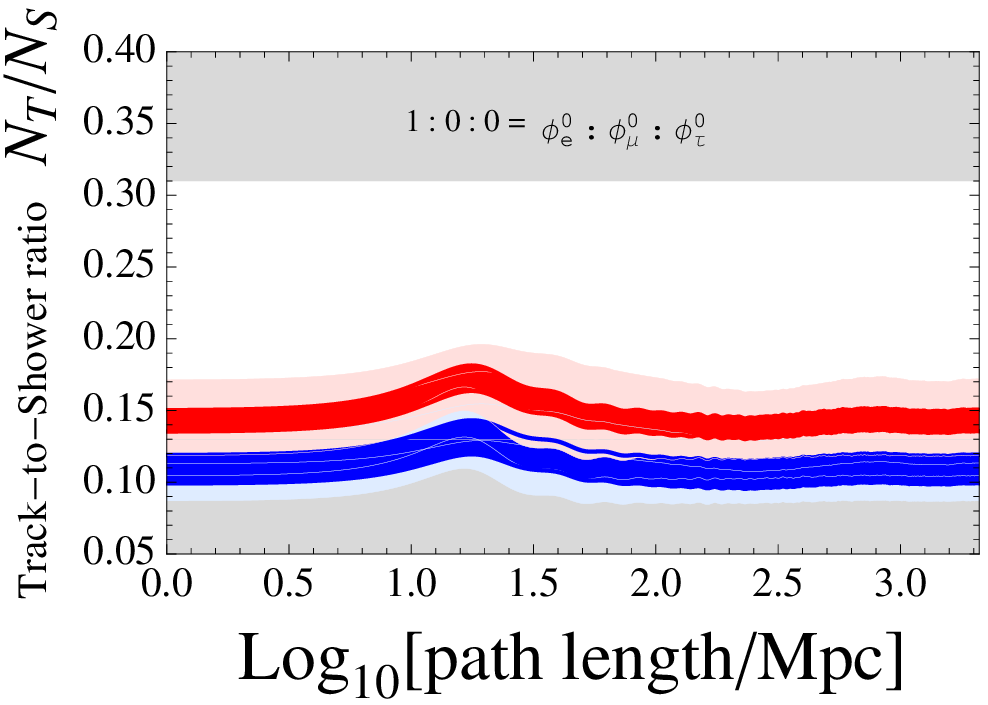,width=8.5cm,angle=0}
\end{minipage}
\hspace*{1.0cm}
\begin{minipage}[h]{7.5cm}
\epsfig{figure=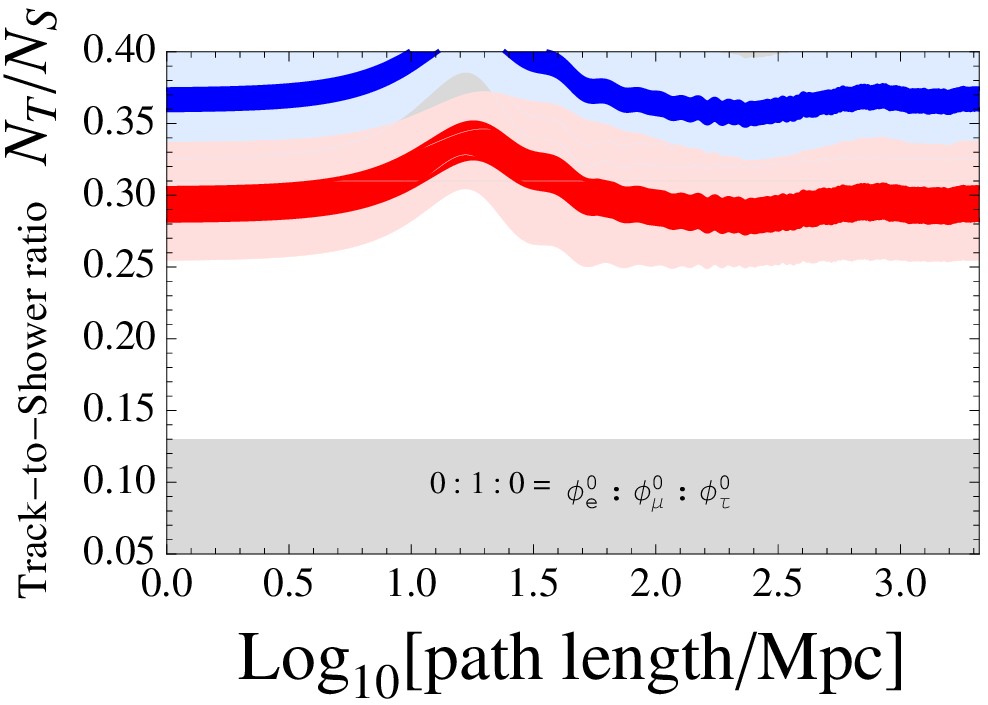,width=8.5cm,angle=0}
\end{minipage}
\caption{\label{FigA2} Plots of the track-to-shower ratio $N_{T}/N_S$ for normal (inverted) mass ordering as a function of $L~(\log_{10}[{\rm path~length/Mpc}])$ for $\Delta m^{2}_{1,2,3}=10^{-16}\,{\rm eV}^2$ .
Each panel corresponds to the specific initial flavor composition ($\phi^{0}_e: \phi^0_\mu: \phi^0_\tau$) at the source.
  For three neutrino mixing angles and Dirac-type CP phase, we take the global fit results at $1\sigma$ ~\cite{Gonzalez-Garcia:2015qrr}.
Red and blue bands correspond to normal and inverted neutrino mass orderings, respectively, for $\alpha=2.2$, whereas light red and light blue regions represent the corresponding results for $\alpha=1.8-2.6$.
Gray shaded regions represent the forbidden bound from $N_T/N_S=0.18^{+0.13}_{-0.05}$ in Ref.~\cite{Palladino:2015zua}.}
\end{figure}

Our numerical results depend on the initial flavor composition $\phi^0_e:\phi^0_\mu:\phi^0_\tau$ at the source which are relevant for the interpretation of observational data.
We consider the well-known four production mechanisms for high energy neutrinos from which the flavor compositions are given as : (i) $(\frac{1}{3}: \frac{2}{3}: 0)$ for $\pi$ decay, (ii) $(\frac{1}{2}: \frac{1}{2}: 0)$ for charmed mesons decay, (iii) $(1: 0 : 0)$ for $\beta$ decay of neutrons, and (iv) $(0 : 1 : 0)$ for $\pi$ decay with damped muons.
The tiny mass splittings $\Delta m^2_k$ can be searched for, looking at high energy cosmic neutrinos by measuring the track-to-shower ratio $N_T/N_S$ as the function of $L~(\log_{10}[{\rm path~length}/{\rm Mpc}])$ in Eq.~(\ref{NTS}).
In the numerical analysis, we use the spectral index given by $\alpha=2.2\pm0.4$~\cite{Aartsen:2013jdh}.

\subsection{Results for the case of $\Delta m^2_{1,2,3}=\Delta m^2_{k}$}
 As a benchmark point, we take the mass splittings $\Delta m^{2}_{1,2,3}$ to be $10^{-16}\,{\rm eV}^2$.
In Fig.~\ref{FigA2}, we present the track-to-shower ratio $N_{T}/N_S$ for normal (inverted) mass ordering as a function of $L~(\log_{10}[{\rm path~length}/{\rm Mpc}])$.
The red (dark black) and blue (light black) curves correspond to normal and inverted neutrino mass orderings, respectively, for $\alpha=2.2$.
The light red and light blue regions correspond to normal and inverted neutrino mass orderings, respectively, for $1.8\lesssim \alpha\lesssim 2.6$.
The gray shaded regions are forbidden by the measurement of $N_T/N_S=0.18^{+0.13}_{-0.05}$ which is obtained in Ref.~\cite{Palladino:2015zua}.
In each panel, we present the initial flavor composition for neutrino flux.
The width of each band in the panels represent the uncertainties in the measurements of neutrino mixing angles. From Fig.~\ref{FigA2}, we see that the bands for the cases with only one flavor in the initial flavor composition are wider than the others. This is because of the slightly hierarchical neutrino mixing angles and the initial flavor compositions with $\phi_\ell$ in Eq.~(\ref{NTS}): for example, the main reason in the different band widths of the left upper and lower panels in Fig.~\ref{FigA2} is the initial flavor compositions with $\phi_\ell$, while the main reason in the different band widths of the left-lower and right-lower panels is the slightly hierarchical neutrino mixing angles.
The results in the lower panels show that the predicted values of $N_T/N_S$ for normal mass ordering are consistent with the boundaries of the allowed region of $N_T/N_S$.
In the upper panels, we see that the oscillation peak occurs at the distance about $1.3$ Gpc for both the tiny mass splittings $\Delta m^{2}_{1,2,3}=10^{-16}\,{\rm eV}^2$ and the deposited energies $60\,{\rm TeV}-3\,{\rm PeV}$.

\subsection{Results for the case of $\Delta m^2_{i}\gg\Delta m^2_{j}$}
\begin{figure}[h]
\begin{minipage}[h]{7.5cm}
\epsfig{figure=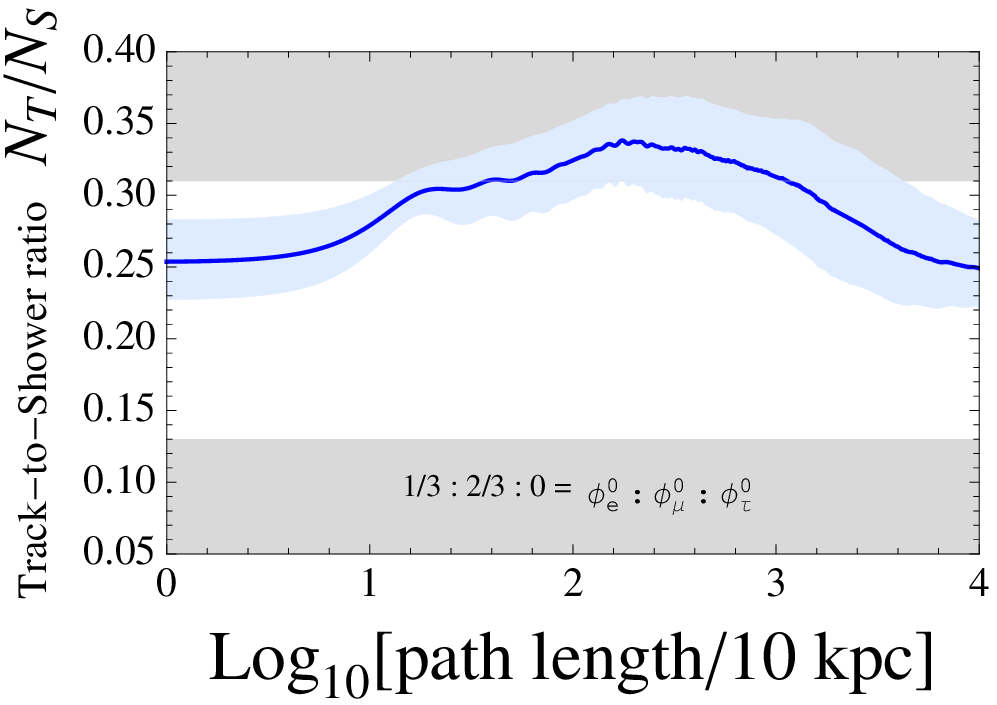,width=8.5cm,angle=0}
\end{minipage}
\hspace*{1.0cm}
\begin{minipage}[h]{7.5cm}
\epsfig{figure=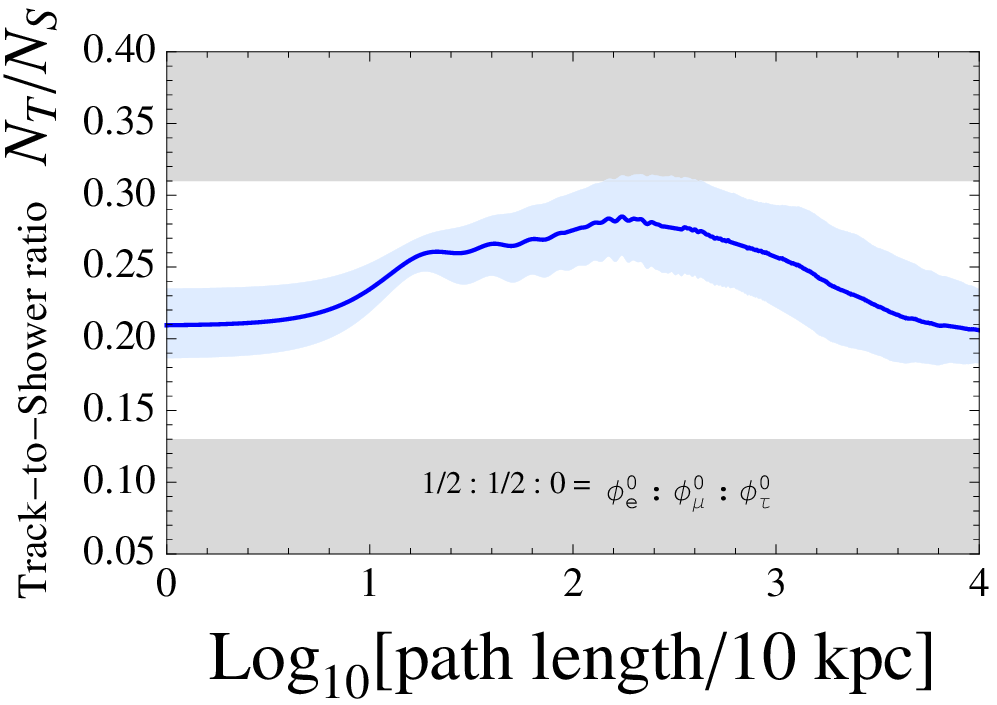,width=8.5cm,angle=0}
\end{minipage}\\
\begin{minipage}[h]{7.5cm}
\epsfig{figure=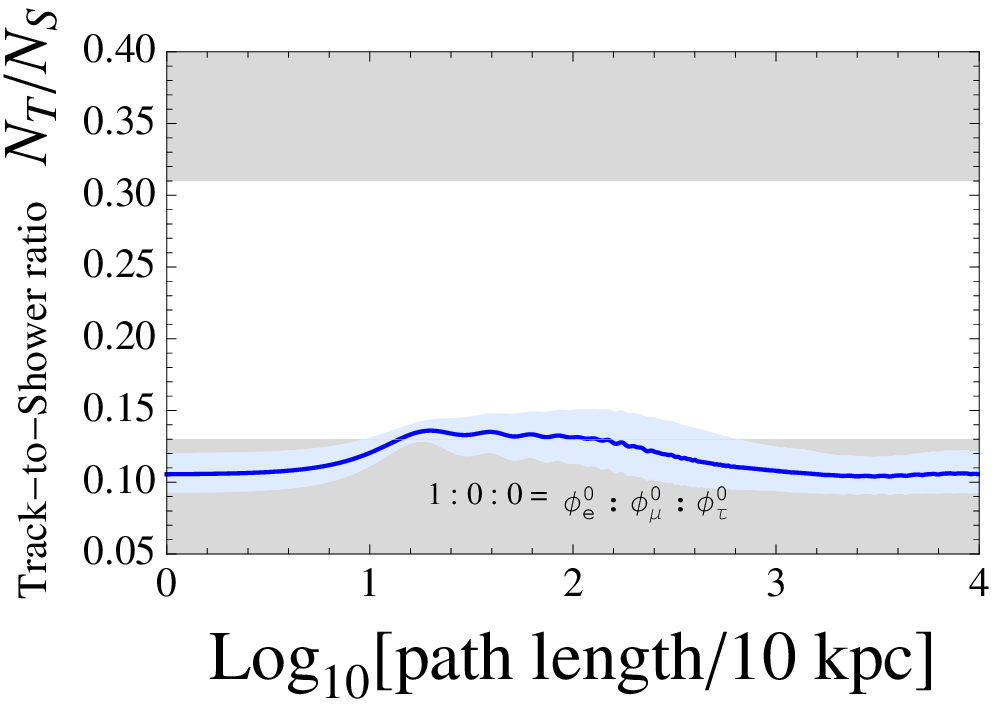,width=8.5cm,angle=0}
\end{minipage}
\hspace*{1.0cm}
\begin{minipage}[h]{7.5cm}
\epsfig{figure=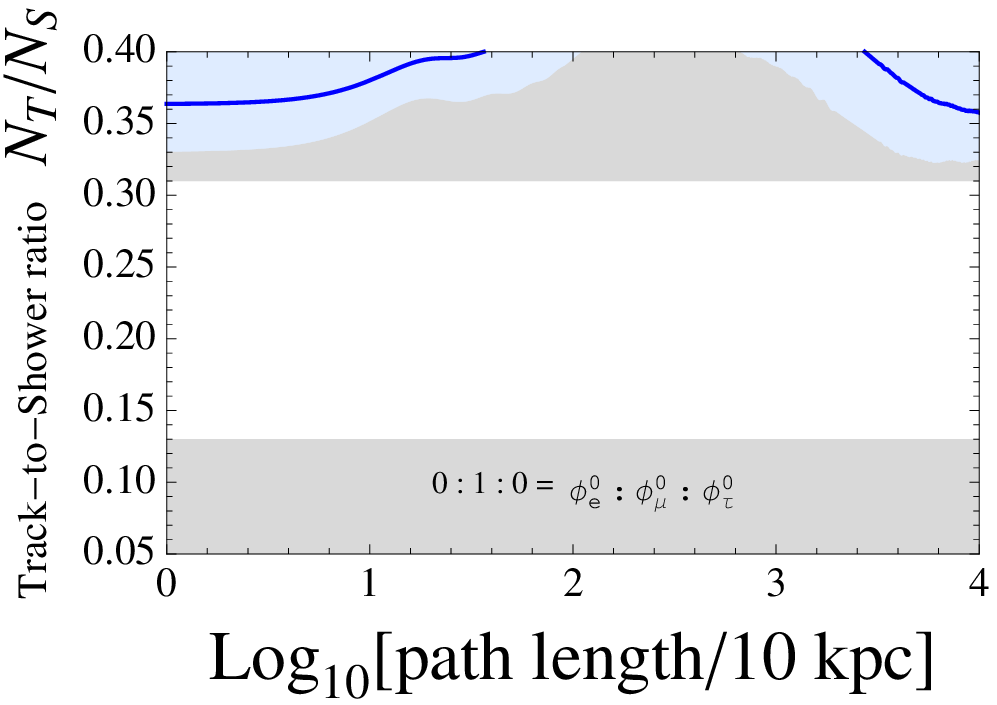,width=8.5cm,angle=0}
\end{minipage}
\caption{\label{FigA3}
Plots of $N_{T}/N_S$ for inverted neutrino mass ordering as a function of $L~(\log_{10}[{\rm path~length/Mpc}])$ for  $\Delta m^{2}_{1}=10^{-14}\,{\rm eV}^2$, $\Delta m^{2}_{2}=10^{-15}\,{\rm eV}^2$, and $\Delta m^{2}_{3}=10^{-16}\,{\rm eV}^2$.
The input values except for $ \Delta m^{2}_{k}$ are taken to be the same as in Fig.~\ref{FigA2}.
Gray shaded regions represent the forbidden bound from $N_T/N_S=0.18^{+0.13}_{-0.05}$ in Ref.~\cite{Palladino:2015zua}.
Each panel corresponds to the specific initial flavor composition at the source, as in Fig. 2.}
\end{figure}

\begin{figure}[h]
\begin{minipage}[h]{7.5cm}
\epsfig{figure=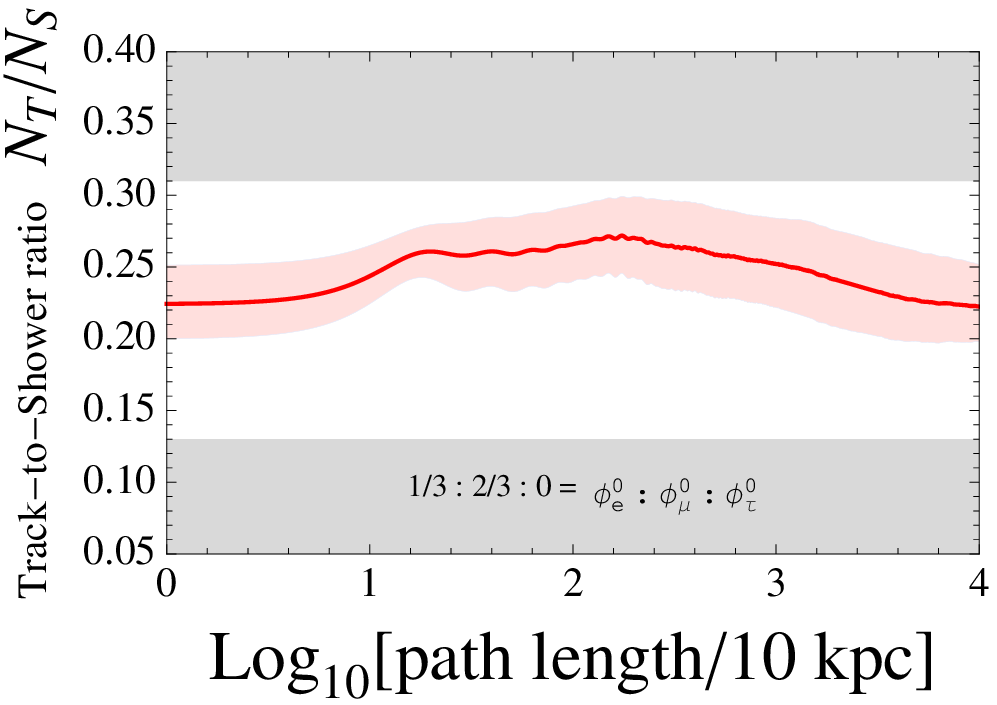,width=8.5cm,angle=0}
\end{minipage}
\hspace*{1.0cm}
\begin{minipage}[h]{7.5cm}
\epsfig{figure=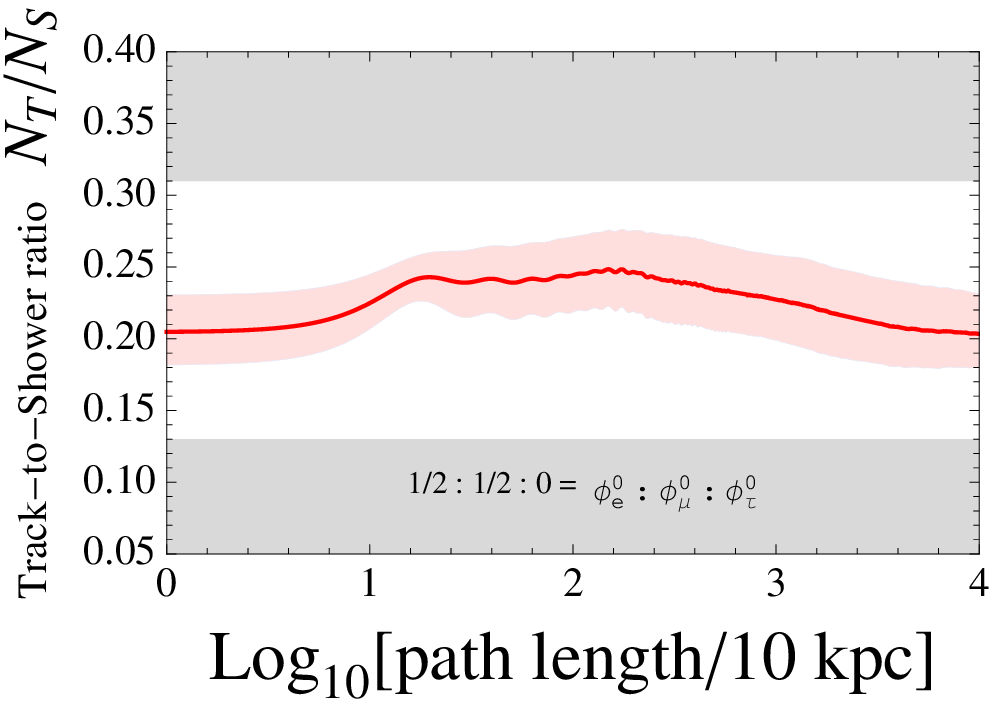,width=8.5cm,angle=0}
\end{minipage}\\
\begin{minipage}[h]{7.5cm}
\epsfig{figure=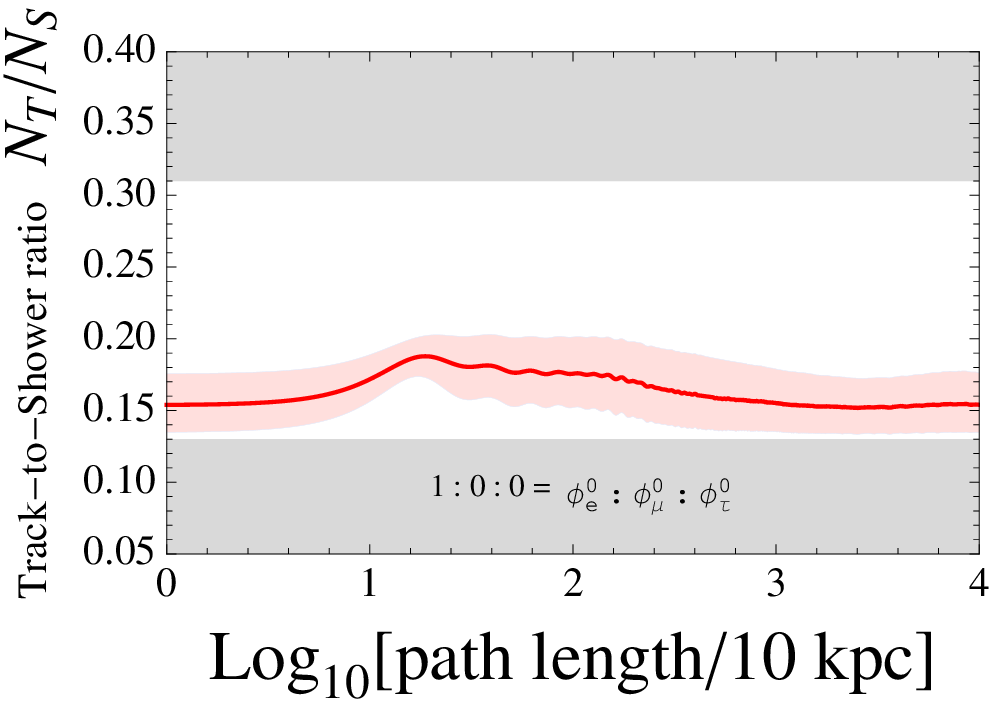,width=8.5cm,angle=0}
\end{minipage}
\hspace*{1.0cm}
\begin{minipage}[h]{7.5cm}
\epsfig{figure=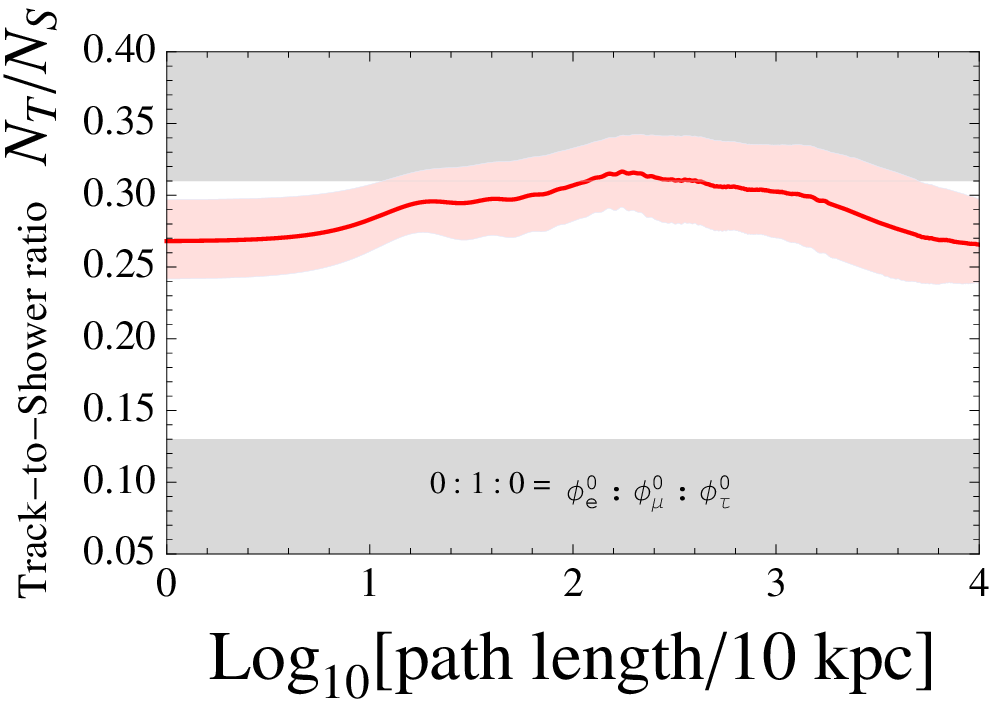,width=8.5cm,angle=0}
\end{minipage}
\caption{\label{FigA4}
Plots of $N_{T}/N_S$ for normal neutrino mass ordering as a function of $L~(\log_{10}[{\rm path~length/Mpc}])$ for  $\Delta m^{2}_{1}=10^{-14}\,{\rm eV}^2$, $\Delta m^{2}_{2}=10^{-15}\,{\rm eV}^2$, and $\Delta m^{2}_{3}=10^{-16}\,{\rm eV}^2$.
The input values except for $\Delta m^2_{k}$ are taken to be the same as in~\ref{FigA2}.
Gray shaded regions represent the forbidden bound from $N_T/N_S=0.18^{+0.13}_{-0.05}$ in Ref.~\cite{Palladino:2015zua}.
Each panel corresponds to the specific initial flavor composition at the source, as in Fig. 2.}
\end{figure}

In this case, we take $\Delta m^{2}_{1}=10^{-14}\,{\rm eV}^2$, $\Delta m^{2}_{2}=10^{-15}\,{\rm eV}^2$, and $\Delta m^{2}_{3}=10^{-16}\,{\rm eV}^2$ as a benchmark point.
In the numerical analysis, we take the input values except for $\Delta m^2_{k}$ to be the same as in Fig.~\ref{FigA2}.
In Figs.~\ref{FigA3} and \ref{FigA4}, we plot the track-to-shower ratio $N_{T}/N_S$ as a function of $L~(\log_{10}[{\rm path~length/Mpc}])$
for the inverted and normal neutrino mass orderings, respectively.
Initial flavor compositions at the source are the same as in Fig.~\ref{FigA2}.
Gray shaded regions represent the forbidden bound from $N_T/N_S=0.18^{+0.13}_{-0.05}$ in Ref.~\cite{Palladino:2015zua}.
We can see from Fig.~\ref{FigA3} that the results for the case with initial flavor composition $(\phi^0_e:\phi^0_\mu:\phi^0_\tau)=(0 : 1 : 0)$ at the source is not
consistent with experimental results, whereas only the highest region for the case with  $(\phi^0_e:\phi^0_\mu:\phi^0_\tau)=(1 : 0 : 0)$
is consistent with experimental results.
As can be seen from ~Fig.\ref{FigA4}, the predictions of $N_T/N_S$ for the normal mass ordering  are all consistent with experimental results.
Different from the results of Fig.~\ref{FigA2}, the oscillation peaks in each panels in Figs.~\ref{FigA3} and \ref{FigA4} are not so sharp.
This means that the predictions of $N_T/N_S$ in these cases are deviated from the case with no new oscillatory effect for rather wider regions of the parameter $L$.

Up to this point, we have presented the numerical results for $N_T/N_S$ as a function of $L$ for a given $\Delta m^2_{k}$ and
$60~{\rm TeV}\lesssim E \lesssim 3~{\rm PeV}$.
However, the experimental results on high energy neutrinos released from IceCube are shown 
for narrow interval of energy from 60 TeV to 3 PeV.
From our numerical results, we see that the amount to which the prediction for $N_T/N_S$ is deviated from the one without new oscillatory effect
depends on the energy scale.
In Table \ref{tab2}, we present how the prediction for $N_T/N_S$ is deviated from the case without new oscillatory effect for $\Delta m^2_{k}=10^{-16}~ {\rm eV}^2$
and $L(z)=1.1$ Gpc.
According to Table \ref{tab2}, we see that the largest deviation occurs for $2.286 \lesssim E \lesssim 3$ PeV.
The deviation becomes smaller as $E$ goes lower. Similar result is obtained for the hierarchical case.
Thus, more PeV scale data would be desirable to test our model.

\begin{table}
\caption{Deviation of the prediction for $N_T/N_S$ from the case without new oscillatory effect for $\Delta m^2_{k}=10^{-16}~ {\rm eV}^2$.}
\centering
\begin{tabular}{c||c|c|c|c|c|c|c}
\hline \hline
$E$ (PeV)&(0.523,0.627)&(0.627,0.754)&(0.754,0.905)&(0.905,1.312)&(1.312,1.576)&(1.576,2,286)& (2.286,3) \\  \hline
Dev($\%$)&15.7&15.9&16.0&16.1&16.2&16.3&16.4 \\
\hline
\end{tabular}
\label{tab2}
\end{table}

In order to probe the presence of peudo-Dirac neutrino, observation of new oscillatory effects in $N_T/N_S$ is essential.
To do this, future experiments should precisely measure the value of $N_T/N_S$.
If the uncertainty in future measurements could be reduced by 50-60 $\%$ from the current one without changing the central value, there would be a high chance to observe the new oscillatory effects via the oscillation peak for the case of degenerate $\Delta m^2_{k}$, and we would be able to
test the pseudo-Dirac property of neutrinos,
particularly for the cases with two flavors in the initial flavor composition of neutrino flux.
For the case of hierarchical $\Delta m^2_{k}$, to test the model for peudo-Dirac neutrino, we need to reduce the uncertainty by 40-70 $\%$ depending
on the initial flavor compositions.

As expected, for $3\sigma$ data of three neutrino oscillations one could not distinguish normal and inverted orderings for the track-to-shower ratio $N_T/N_S$, while the band width can be enlarged.

We can constrain mass squared differences $\Delta m^2_1$ and $\Delta m^2_2$
from the fact that the UHE neutrinos with energy $10^{9}$ GeV are expected from the Greisen-Zatsepin-Kuzmin limit cosmic rays originated at distances of $100$ Mpc~\cite{PAC}.
To observe such UHE neutrinos through neutrino oscillation whose length is of order $100$ Mpc, the required magnitudes of  $\delta_1$ and $\delta_2$ can be estimated from
\begin{eqnarray}
 L^{1,2}_{\rm osc}\simeq\left(\frac{0.8\times10^{-12}\,\text{eV}^2}{\Delta m^{2}_{1,2}}\right)\left(\frac{E}{10^9\,\text{GeV}}\right)100\,\text{Mpc}\lesssim100\,\text{Mpc}\,,
\end{eqnarray}
which means that for such neutrinos with $E\sim10^9$ GeV, oscillation length will be order of 100 Mpc for $\Delta m^{2}_{1,2}\simeq10^{-12}\,\text{eV}^2$. In other words, $\delta_{1,2}\geq4.6\times10^{-11}$ eV (for normal mass hierarchy) and $\delta_{1,2}\geq0.8\times10^{-11}$ eV (for inverted mass hierarchy) are required for significant conversion of these neutrinos. Taking into account oscillation length of order the earth-sun distance 1A.U for  neutrino energy $10^9$ GeV, we estimate the mass splittings are so large $\Delta m^2_{1,2}\simeq16.6\,{\rm eV}^2$ which contradicts with $\Delta m^2_k\ll \Delta m^2_{\rm sol}$.
Thereby, electron neutrinos from the nearby sources and high energy can remain undepleted, but ones from extragalactic sources get depleted.


\section{Conclusion}

In this work, we have proposed a model where sterile neutrinos are introduced to make light neutrinos  to be pseudo-Dirac particles.
It has been shown how tiny mass splitting necessary for realizing pseudo-Dirac neutrinos can be achieved.
Within the model, we have examined how leptogenesis can be successfully generated.
Motivated by the recent observation of  very high energy neutrino events at IceCube and the results for  the track-to-shower ratio, $N_T/N_S$, of
the subset with energy above 60 TeV studied in Ref. \cite{Palladino:2015zua},
we have examined a possibility to observe the effects of the pseudo-Dirac property of neutrinos
by performing astronomical-scale baseline experiments to uncover the oscillation effects of very tiny mass splitting.
Using  the result of global fit to neutrino data for the input of neutrino mixing angles and CP phase at $1\sigma$ C.L. and fixing  neutrino energy and mass splittings,
we have studied how the oscillation effects induced by pseudo-Dirac neutrinos may affect the track-to-shower ratio, and found
that the oscillation peaks occur at the distance about 1.3 Gpc for $\Delta m^2_{1,2,3}=10^{-16}\,\mbox{eV}^2$ and $60\,\mbox{TeV}\lesssim E_{\nu} \lesssim 3\,\mbox{PeV}$. 
If future experiments can precisely measure the value
of $N_T/N_S$, whose uncertainty becomes reduced to about $40-70\%$ depending on the initial flavor compositions at the source,  we could test the pseudo-Dirac property of neutrinos
particulary for the cases with two flavors in the initial flavor composition of neutrino flux.
In fact, in order to obtain fully meaningful results for testing our model in detail,  much larger detectors than the present IceCube would be required \cite{larger}.

\begin{acknowledgements}
We woule like to thank Francis Halzen for valuable comments on the ultra-high energy cosmic neutrinos.
The work of SKK is supported by NRF-2014R1A1A2057665. The work of Y.H. Ahn is supported by IBS under the project code, IBS-R018-D1. The work of CSK was supported by NRF grant funded by the Korea government of the MEST (No. 2011-0017430), (No. 2011-0020333).

\end{acknowledgements}


\end{document}